# Magnetization Reversal Signatures of Hybrid and Pure Néel Skyrmions in Thin Film Multilayers

Nghiep Khoan Duong[1]*, Riccardo Tomasello[2]*, M. Raju[1], Alexander P. Petrović[1], Stefano Chiappini[3], Giovanni Finocchio[3,4]^, Christos Panagopoulos[1]^

[1]Division of Physics and Applied Physics, School of Physical and Mathematical Sciences, Nanyang Technological University, S637371, Singapore
[2]Institute of Applied and Computational Mathematics, Foundation for Research and Technology – Hellas (FORTH), GR-70013 Heraklion-Crete, Greece
[3]Istituto Nazionale di Geofisica e Vulcanologia, Via di Vigna Murata 605, I-00143 Roma, Italy
[4]Department of Mathematical and Computer Sciences, Physical Sciences and Earth Sciences, University of Messina, I-98166 Messina, Italy

Abstract

We report a study of the magnetization reversals and skyrmion configurations in two systems – Pt/Co/MgO and Ir/Fe/Co/Pt multilayers, where magnetic skyrmions are stabilized by a combination of dipolar and Dzyaloshinskii-Moriya interactions (DMI). First Order Reversal Curve (FORC) diagrams of low-DMI Pt/Co/MgO and high-DMI Ir/Fe/Co/Pt exhibit stark differences, which are identified by micromagnetic simulations to be indicative of hybrid and pure Néel skyrmions, respectively. Tracking the evolution of FORC features in multilayers with dipolar interactions and DMI, we find that the negative FORC valley, typically accompanying the positive FORC peak near saturation, disappears under both reduced dipolar interactions and enhanced DMI. As these conditions favor the formation of pure Neel skyrmions, we propose that the resultant FORC feature – a single positive FORC peak near saturation – can act as a fingerprint for pure Néel skyrmions in multilayers. Our study thus expands on the utility of FORC analysis as a tool for characterizing spin topology in multilayer thin films.

*These authors contributed equally to this work

^Corresponding authors: gfinocchio@unime.it, christos@ntu.edu.sg

I. Introduction

Magnetic skyrmions have been realized in several material systems, most notably magnetic multilayer thin films which host nanoscale skyrmions at room temperature[1-3]. In such multilayers, the Dzyaloshinskii-Moriya interaction (DMI) arising at the ferromagnet (FM)/heavy-metal (HM) interfaces is paramount in stabilizing the Néel spin textures that these skyrmions possess[4, 5]. However, the actual spin textures of these skyrmions have recently been proven to be more complex, owing to the competition between DMI and dipolar interactions between the thin film layers[6-8]. For most multilayer systems, skyrmions exhibit thickness-dependent magnetization profiles, where a central-layer Bloch texture is sandwiched between Néel textures of opposite chiralities from the topmost and bottommost layers[7]. These are known as hybrid skyrmions, whereas uniform Néel-texture skyrmions throughout the multilayer, realizable in a high-DMI environment, are known as pure Néel skyrmions[7].

Differences in the spin texture and chirality of skyrmions strongly influence their current-driven dynamics[7, 9-11], rendering the knowledge of their complete, three-dimensional spin textures crucial for spintronic material design. Distinguishing between hybrid and pure Néel skyrmions, however, requires sophisticated imaging methods, such as circular dichroism x-ray resonant magnetic scattering[7, 8] or nitrogen-vacancy center magnetometry[12] in order to resolve the thickness-dependence of the magnetic textures. These techniques may not always be readily available in most research facilities. On the other hand, the interplay of dipolar interactions and DMI, as well as other ubiquitous and tunable magnetic interactions in multilayers, directly affect the domain size, density, and the level of disorder in the skyrmion configuration[3]. These parameters consequently influence the magnetization reversal processes[13] and hysteretic behaviour of spin textures[14], which may provide indirect clues for inferring the inner complexity of three-dimensional skyrmions.

The ability to identify these subtle processes has been demonstrated by the First Order Reversal Curve (FORC) technique, which provides a magnetic fingerprint of the interactions and reversal processes occuring in magnetic materials[15, 16]. Recent studies have begun utilizing FORC in skyrmion-hosting multilayers to study field history control of zero-field skyrmion population[17, 18], while simultaneously revealing magnetic reversal mechanisms influenced by the skyrmion configuration[17]. Indeed, the variety of magnetic interactions and skyrmion configurations realizable in different thin film heterostructures offer a rich resource for FORC studies.

Making use of this approach, our work combines FORC and magnetic force microscopy (MFM) to study the magnetization reversal and skyrmion properties in two systems: [Pt(3)/Co(0.9)/MgO(1.5)] x15 and [Ir(1)/Fe(0.4)/Co(0.4)/Pt(1)]x20 (thickness in nm in parentheses), hereafter referred to as Pt/Co/MgO and Fe(0.4)/Co(0.4), respectively. The former displays skyrmions with a large average diameter of $105 \pm 21\ nm$, suggesting that their stability results primarily from magnetic dipolar interactions[19], while the latter shows smaller $46 \pm 12\ nm$ skyrmions, indicating the more dominant role played by the DMI (*D*) in skyrmion formation[19]. FORC analysis and MFM imaging reveal distinct irreversibility features in these two material systems. Using micromagnetic simulations, we show these two multilayers stabilize hybrid and pure Néel skyrmions, respectively, which may account for their distinct FORC features. To support this hypothesis, we apply our analysis to Pt/Co/MgO samples with different numbers of layer repetitions, and also to Fe/Co multilayers with different ferromagnetic compositions. Again, we observe a correlation between FORC features and the relative strengths of dipolar interactions and DMI, which faciliate the transition from a hybrid to a pure Néel skyrmion texture [7]. This points towards a possible thermodynamic signature for high-*D* multilayers, which can stabilize pure Néel skyrmions.

II. Methods

Multilayer thin films of Ta(3)/Pt(10)/[Ir(1)/Fe(0.2-0.4)/Co(0.4-0.6)/Pt(1)]$_{20}$/Pt(2) and Ta(3)/[Pt(3)/Co(0.9)/MgO(1.5)]$_{15}$/Ta(4) were deposited on thermally oxidized silicon wafers at room temperature (numbers in parentheses refer to the layer thickness in nm). Ir/Fe/Co/Pt samples were deposited using a Chiron ultra-high vacuum multi-source sputter tool, while Pt/Co/MgO samples were deposited using a Singulus Timaris ultra-high vacuum multi-target sputter tool. The base vacuum in each case is 1×10$^{-8}$ Torr and sputtering is carried out in 1.5×10$^{-3}$ Torr of argon gas. Magnetization measurements on these samples were performed using superconducting quantum interference device (SQUID) magnetometry, in a Quantum Design Magnetic Property Measurement System (MPMS), to obtain the saturation magnetization ($M_S$). Out-of-plane and in-plane hysteresis loops were also acquired to determine the uniaxial effective anisotropy values $K_{eff}$.

FORC measurements were then conducted on as-grown samples using a Vibrating Sample Magnetometer (VSM) at room temperature. Each FORC measurement consists of a two-part sequence: (1) the sample is first saturated at a positive field and then brought to a reversal field $H_R$, (2) from $H_R$ the magnetization of the sample is measured starting from $H_R$ and ending at 0, as the applied field is reversed. Repeating the sequence for multiple values of $H_R$, we obtain a family of FORCs (Figure 1(a),(b)), used to compute the FORC distribution $\rho$ defined as:

$$\rho = -\frac{1}{2}\frac{\partial^2 M(H_R,H)}{\partial H_R \partial H} \quad (1)$$

Plotting $\rho$ as a density plot against $H_R$ and $H$ produces a FORC diagram (Figure 1(c), (g)), which quantifies the degree of magnetic irreversibility for the magnetic field histories of the measured sample. Each FORC diagram is studied by complementary MFM images, which capture the magnetic textures obtained by different field histories. The method used for acquiring and analyzing MFM images is similar to that described in Ref. [3].

Micromagnetic computations were performed by means of a state-of-the-art micromagnetic solver, PETASPIN, which numerically integrates the Landau-Lifshitz-Gilbert (LLG) equation by applying the Adams-Bashforth time solver scheme[20]:

$$\frac{d\mathbf{m}}{d\tau} = -(\mathbf{m}\times\mathbf{h}_{eff}) + \alpha_G\left(\mathbf{m}\times\frac{d\mathbf{m}}{d\tau}\right), \quad (2)$$

where $\alpha_G$ is the Gilbert damping, $\mathbf{m} = \mathbf{M}/M_s$ is the normalized magnetization, and $\tau = \gamma_0 M_s t$ is the dimensionless time, with $\gamma_0$ being the gyromagnetic ratio and $M_s$ the saturation magnetization. $\mathbf{h}_{eff}$ is the normalized effective magnetic field, which includes the exchange, interfacial DMI, uniaxial anisotropy, and Zeeman fields, as well as the magnetostatic field computed by solving the magnetostatic problem of the whole system [8, 21].

The [Pt(3 nm)/Co(0.9 nm)/MgO(1.5 nm)]$_{15}$ and [Ir(1 nm)/Fe(0.4 nm)/Co(0.4 nm)/Pt(1 nm)]$_{20}$ multilayers are simulated by 15 and 20 repetitions of a 1 nm FM, respectively. The ferromagnetic layers are coupled to one another by means of only the magnetostatic field (exchange-decoupled layers), and are separated by a non-magnetic layer: 4 nm thick Pt/MgO for Pt/Co/MgO, and 2 nm thick Ir/Pt for Fe(0.4)/Co(0.4). The ferromagnetic and spacer thicknesses have been chosen to reduce the number of cells along the z-direction, while matching the experimental layer thicknesses as closely as possible. For Pt/Co/MgO, the physical parameters used are as follows: saturation magnetization $M_s$=1.586 MA/m, $K_{eff}$=49.56 kJ/m$^3$ (determined by SQUID measurements) exchange constant $A$=12 pJ/m (based on Refs. [2, 7]), interfacial $D$=0.5 mJ/m$^2$ (calculated from the periodicity of MFM-imaged domain walls, based on the method described in Ref [22]), and a discretization cell size of 2.5x2.5x1 nm$^3$. An out-of-plane external field $H_{ext}$ = 110 mT is applied antiparallel to the skyrmion core, as in the MFM measurements. For Fe(0.4)/Co(0.4), we used $M_s$=841 kA/m, $K_{eff}$=109.38 kJ/m$^3$, $A$=11 pJ/m, $D$=2.1 mJ/m$^2$ (according to Ref. [3]), $H_{ext}$ = 160 mT (as in the MFM measurements), and a discretization cell size 3x3x1 nm$^3$.

    III.    Results and discussion

We first focus on the FORC diagram of Pt/Co/MgO, which has an estimated (Refs. [2, 7]) $D$ value of 0.5 mJ/m$^2$. For this sample, we observe large regions of irreversibility extending all the way from $H_R = 0$ mT to $H_R \approx -180$ mT. The first feature is a wide, positive-valued ridge from $\mu_0 H_R = 0\ mT$ to $\mu_0 H_R \approx -125\ mT$ (Figure 1(c)), coinciding with the transition from labyrinthine stripes to the skyrmion phase (Figure 1(d)-(f)), where approximately 100 nm-diameter skyrmions emerge in a disordered configuration at $\mu_0 H = \mu_0 H_R = -110\ mT$ (Figure 1(f)). Based on the interpretation of Ref. [17], we deduce that the large, positive-valued region of irreversibility for $|H| \leq |H_R|$ in this $H_R$ range corresponds to skyrmion and stripe mergers taking place as the applied field decreases. As $|\mu_0 H_R|$ increases from $125\ mT$ to $180\ mT$, a pair of irreversible regions consisting of a negative-valley (blue) and positive-peak (red) emerges (Figure 1(c)). This familiar pair feature arises from the sign change in the second derivative of the magnetization as neighboring reversal curves diverge and then converge in the high field regime (dashed circles in Figure 1(a)). The feature, which was observed near the out-of-plane saturation fields in FORC studies of other magnetic multilayers[15-17], signifies the onset of skyrmion annihilation as the applied field increases along the diagonal edge, followed by skyrmion and stripe nucleation as the field is reduced along the *H* axis.

While the negative-positive pair feature frequently appears in magnetic multilayers, including the Ir/Fe(x)/Co(y)/Pt stacks, it does not appear for Fe(0.4)/Co(0.4), where $D$=2.1 mJ/m$^2$. No sign change in the second derivative of the magnetization is observed, and hence only a single positive peak is seen as the system approaches saturation, i.e. from $\mu_0 H_R \approx -150$ mT to $\mu_0 H_R \approx -225$ mT (Figure 1(g)). This feature is preceded by an elongated irreversible ridge extending from $\mu_0 H_R \approx -50\ mT$ to $\mu_0 H_R \approx -150\ mT$ (Figure 1(g)). Unlike the sprawling irreversible feature in Pt/Co/MgO, the irreversible ridge for Fe(0.4)/Co(0.4) is narrower and localized around the diagonal edge of the FORC diagram. This indicates the presence of a large population of skyrmions, whose repulsive interaction at the short range precludes skyrmion merger, thus resulting in less irreversible activity taking place as the applied field reverses from $H_R$. Indeed, high-density skyrmions appear as early as $\mu_0 H_R \approx -50\ mT$ (Figure 1(h)) and quickly transform into a dense array of small skyrmions ($\approx 50$ nm in diameter) as the applied field increases (Figure 1(i)).

This configuration stands in sharp contrast to the sparse array of larger skyrmions ($\approx 100\ nm$ in diameter) observed in Pt/Co/MgO. Due to their larger size, the latter are likely to be strongly stabilized by dipolar interactions, thus exhibiting hybrid magnetization profiles. The appearance of these hybrid spin textures may be linked to our observed differences in FORC features. To test this hypothesis, we performed micromagnetic simulations of the two systems and extracted their thickness-dependent spin textures.

Figure 2 summarizes micromagnetic simulations for the two multilayers. In both cases, the skyrmion diameter is thickness-dependent, being larger in the middle layer and smaller in the external layers. This is attributed to the *z*-component of the magnetostatic field[7]. The size of the skyrmion is larger in the Pt/Co/MgO sample than in Fe(0.4)/Co(0.4), in qualitative agreement with experimental measurements. A crucial difference between the two cases lies in the thickness-dependence of their respective spin textures. In Fe(0.4)/Co(0.4), the spin chirality is independent of the layer position and a pure Néel skyrmion is obtained in all the layers. This can be attributed to the strong DMI in Fe(0.4)/Co(0.4), which, by overcoming the magnetostatic field dictates the skyrmion texture in all the layers, in agreement with previous theoretical results[7].

On the other hand, a skyrmion in the Pt/Co/MgO exhibits a layer-dependent chirality (hybrid skyrmion), which gradually changes from Néel with an outward spin chirality at the bottom layer, to an intermediate skyrmion mixing Néel-outward and Bloch-clockwise chiralities in the middle layer, and eventually to a Néel skyrmion with inward chirality at the top layer. This is ascribed to the small DMI value in Pt/Co/MgO, thus allowing the magnetostatic field to be dominant. The small DMI only affects the position of the Bloch skyrmion, which is not located in the middle layer, as expected from the magnetostatic field, but is shifted upward to the 10$^{th}$ layer, consistent with previous findings[7, 8].

Comparing our micromagnetic simulations with the FORC features in Fig. 1(c),(g), we found that the coexistence of a positive peak and negative valley of the irreversibility coincides with the stabilization of hybrid skyrmions, stabilized by a combination of DMI and dipolar interactions. On the other hand, the presence of a single positive peak coincides with the presence of pure Néel skyrmions, stabilized primarily by interfacial DMI interactions. The distinct FORC features observed in Fig.1 and the hybrid and pure Néel skyrmion textures suggested by micromagnetic simulations thus suggest a potential correlation between FORC distribution features and the strengths of dipolar interactions and DMI, which influence the thickness-dependent skyrmion textures.

To investigate this correspondence, we track the evolution of FORC distributions and skyrmion diameters in Pt/Co/MgO multilayers with the dipolar interaction strength, by reducing the number of layer repetitions (N) progressively from 15 to 2. The results are encapsulated in Fig 3, where the $H_R$ and $H$ axes of the FORC diagrams are normalized to the out-of-plane saturation field, determined as the $H_R$ value at which irreversible features terminate. As interlayer dipolar interaction weakens with reduced N, the FORC distributions transition from a negative-positive peak pair to a single positive-peak. Correspondingly, the observed skyrmion diameter decreases from $\approx 105$ nm (for N = 15) to $\approx 80$ nm (for N = 4), reflecting a transition from a dipolar-dominant regime to a DMI-dominant regime of skyrmion stability [19]. These observations suggest the disappearance of the negative FORC valley correlates with a reduced dipolar interaction in the multilayer.

Likewise, we also track the evolution of FORC features with the increase in DMI, achieved by varying the Fe/Co compositions of the [Ir(1)/Fe(x)/Co(y)/Pt(1)]$_{20}$ heterostructure. Raising the Fe/Co composition ratio while keeping their total thickness $\leq 1$nm effectively increases the DMI strength while also modifies other magnetic parameters. This results in a variation of the skyrmion size, density, and energetic stability, which can be correlated with key changes in the respective FORC diagrams.

Figures 4 (a)-(e) compare the FORC features for Fe(0.2)/Co(0.6), Fe(0.2)/Co(0.8), Fe(0.3)/Co(0.7), Fe(0.5)/Co(0.5) and Fe(0.4)/Co(0.4). Samples Fe(0.2)/Co(0.6), Fe(0.2)/Co(0.8), and Fe(0.3)/Co(0.7) display the familiar negative-valley/positive-peak feature similar to that exhibited by low-D Pt/Co/MgO (Figure 1(c)). Meanwhile, Fe(0.5)/Co(0.5) shows a large positive peak together with a much smaller negative valley, and Fe(0.4)/Co(0.4) exhibits only a narrow positive-valued FORC ridge. The gradual disappearance of the negative-valley, the increase of the DMI strength, and the two-fold decrease in skyrmion diameter [3, 17](Figure 4 (f)) again suggest a transition from a dipolar-dominant to a DMI-dominant regime of skyrmion stability [19], thus hinting at a transition from hybrid to pure Neel skyrmions.

To support this inference, we have performed additional micromagnetic simulations for samples Fe(0.2)/Co(0.8) and Fe(0.5)/Co(0.5), and compared them with the case of Fe(0.4)/Co(0.4). In Fe(0.2)/Co(0.8) with DMI strength of 1.5 mJ/m$^2$ (Fig. 4(g)), we observe a hybrid skyrmion where the Bloch skyrmion is present in the 17$^{th}$ ferromagnetic layer. In contrast, the Bloch position for Pt/Co/MgO appears roughly at the center of the stack due to dominant dipolar interaction over DMI. In Fe(0.5)/Co(0.5) ($D$ = 1.9 mJ/m$^2$), no Bloch skyrmion is observed, and the 3D skyrmion profile is almost pure-Néel, with outward chirality in all the layers except for the topmost layer, which hosts a Néel skyrmion with inward chirality (Fig. 4(h)). Eventually, the skyrmion profile achieves a complete pure Néel texture in all the layers in the case of Fe(0.4)/Co(0.4), where $D$ reaches 2.0 mJ/m$^2$ (Figure 4(i)). Here we point out that the experimental variation of magnetic interactions and their combined role in the minimization of domain wall energy lead to the stabilization of skyrmion textures with varying spin configurations. The results in Fe/Co multilayers are thus in full agreement with the analysis of the FORC diagrams, and together with the observations in samples with reduced layer repetitions, they suggest the single-positive peak feature in FORC diagram is indicative of the stability of pure Neel skyrmions in thin film multilayers.

    IV.    Conclusion

In summary, we investigated the magnetization reversals and skyrmion configurations for Pt/Co/MgO and Ir/Fe/Co/Pt multilayers, using a combination of FORC measurements, MFM imaging, and micromagnetic simulations. Wide sprawling FORC regions with a characteristic negative-valley/positive-peak pair are indicative of large, hybrid skyrmions in low-*D* Pt/Co/MgO. In contrast, a single positive FORC distribution peak is indicative of small, pure Néel skyrmions in high-*D* Fe(0.4)/Co(0.4). By reducing the number of film layer repetitions in Pt/Co/MgO and tuning the thicknesses of Fe and Co in Fe(x)/Co(y) multilayers, we observe a transition of FORC features from a negative-valley/positive-peak pair to a single positive peak in correspondence with a reduction in dipolar interactions and an increase in the DMI strength, respectively. Hence, we propose that the single positive FORC feature can be a useful fingerprint for pure Néel skyrmions in multilayer systems. In addition to providing an indicator for skyrmion spin chirality, the observed FORC features enable a robust assessment of the thermodynamic stability of skyrmions within a particular multilayer: the negative FORC valley vanishes as the stability rises. Whilst additional spin imaging techniques are desirable for microscopically resolving the multitude of complex spin topologies[7, 8, 12], FORC analysis can play an important role in the analysis of magnetic multilayers. Combining these techniques can efficiently address future challenges in designing and optimizing skyrmionic materials.

Acknowledgements

Work in Singapore was supported by the National Research Foundation (NRF), under NRF Investigatorship programme (Ref. No. NRF-NRFI2015-04) and Ministry of Education (MOE) Singapore, Academic Research Fund (AcRF) Tier 3 (Ref. No. MOE2018-T3-1-002). M.R. thanks the Data Storage Institute of Singapore for sample growth facilities. R.T. and G.F. thank the project "ThunderSKY" funded from the Hellenic Foundation for Research and Innovation and the General Secretariat for Research and Technology, under Grant No. 871.

Data availability statement

The data that support the findings of this study are available from the corresponding author upon reasonable request.

List of Figures:

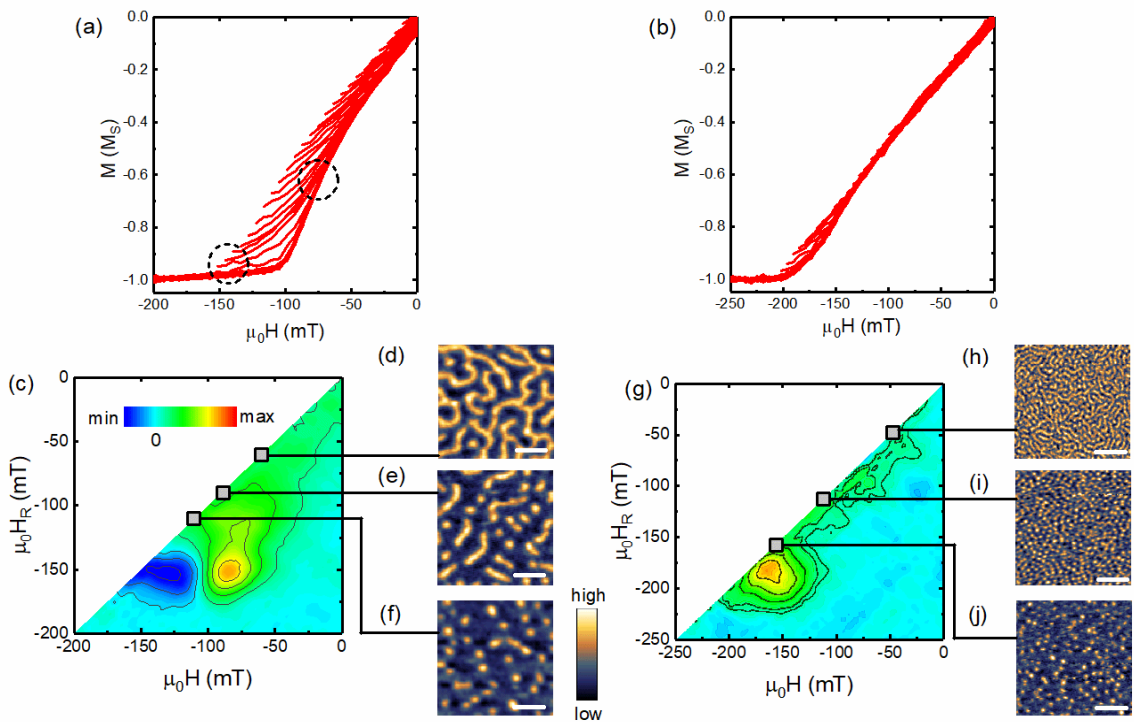

Figure 1: Magnetic irreversibility and skyrmion configurations for low and high DMI systems
A set of FORCs for (a) low-DMI Pt/Co/MgO and (b) high-DMI Fe(0.4)/Co(0.4) multilayers, respectively. Dashed circles in (a) represent points where neighboring reversal curves diverge and converge, resulting in a sign change in the second derivative of the magnetization and consequently leading to the negative-valley/positive-peak pair. (c) FORC diagram of Pt/Co/MgO showing wide-spread irreversible regions terminated with a negative-valley/positive-peak pair near negative saturation. (d)-(f) Field-evolution of spin textures at different $H_R$ values in Pt/Co/MgO. A sparse array of $\approx$ 100 nm skyrmions is seen at $\mu_0 H = \mu_0 H_R = -110$ mT. (g) FORC diagram of Fe(0.4)/Co(0.4) showing narrow irreversible regions constrained to the diagonal edge and terminated with a single positive-peak near negative saturation. (h)-(j) Field evolution of spin textures at different $H_R$ values in Fe(0.4)/Co(0.4). A dense array of $\approx$ 50nm skyrmions is seen at -160 mT in Fe(0.4)/Co(0.4). All magnetic configurations were attained following saturation in a positive magnetic field at T = 300K. Scale bars: 500 nm.

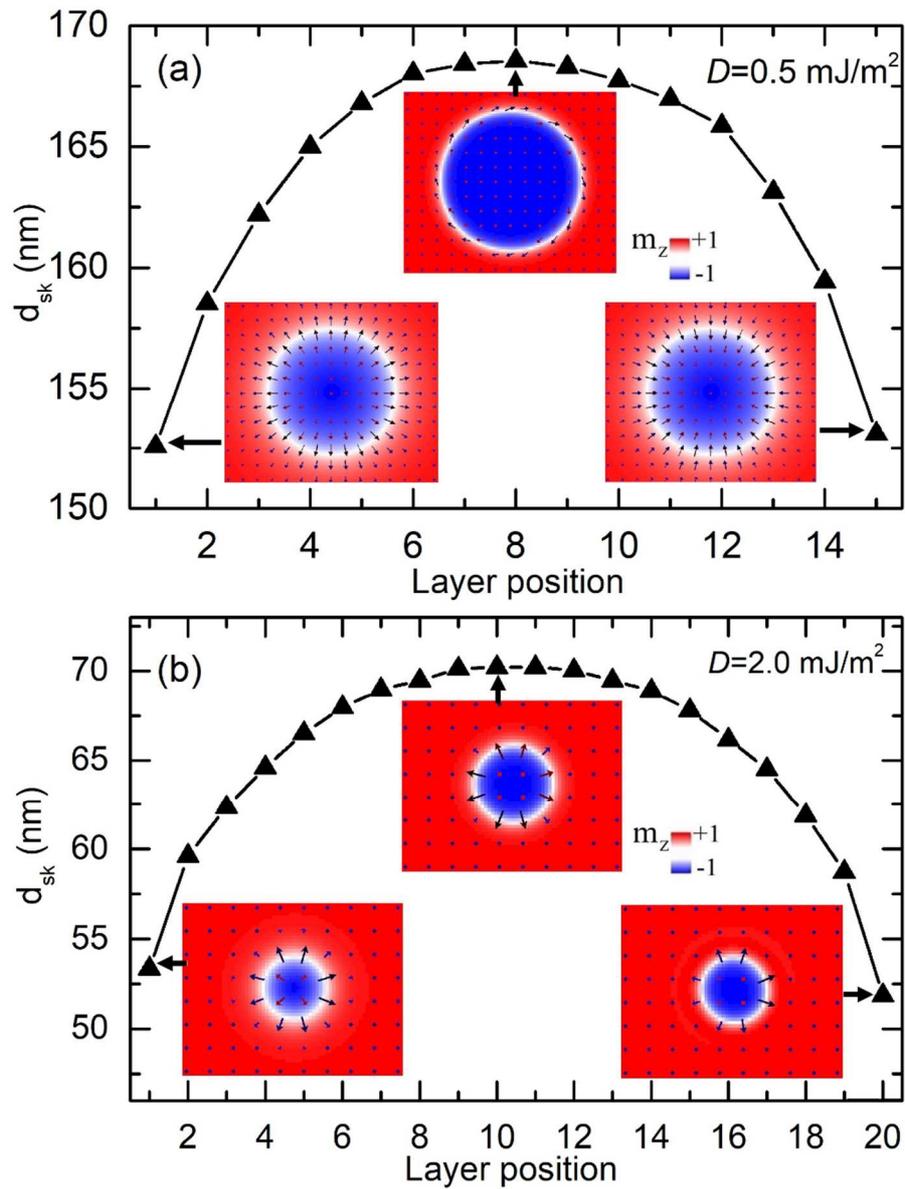

Figure 2: Micromagnetic simulations of skyrmion textures in low and high DMI systems

Skyrmion diameter as a function of layer position for (a) Pt/Co/MgO at $H_{ext}$=110 mT and (b) Fe(0.4)/Co(0.4) at $H_{ext}$=160 mT. The insets depict the spatial distribution of the skyrmion magnetization at the layer positions indicated by the black arrows. The colors correspond to the normalized z-component of the magnetization, as shown in the color bar.

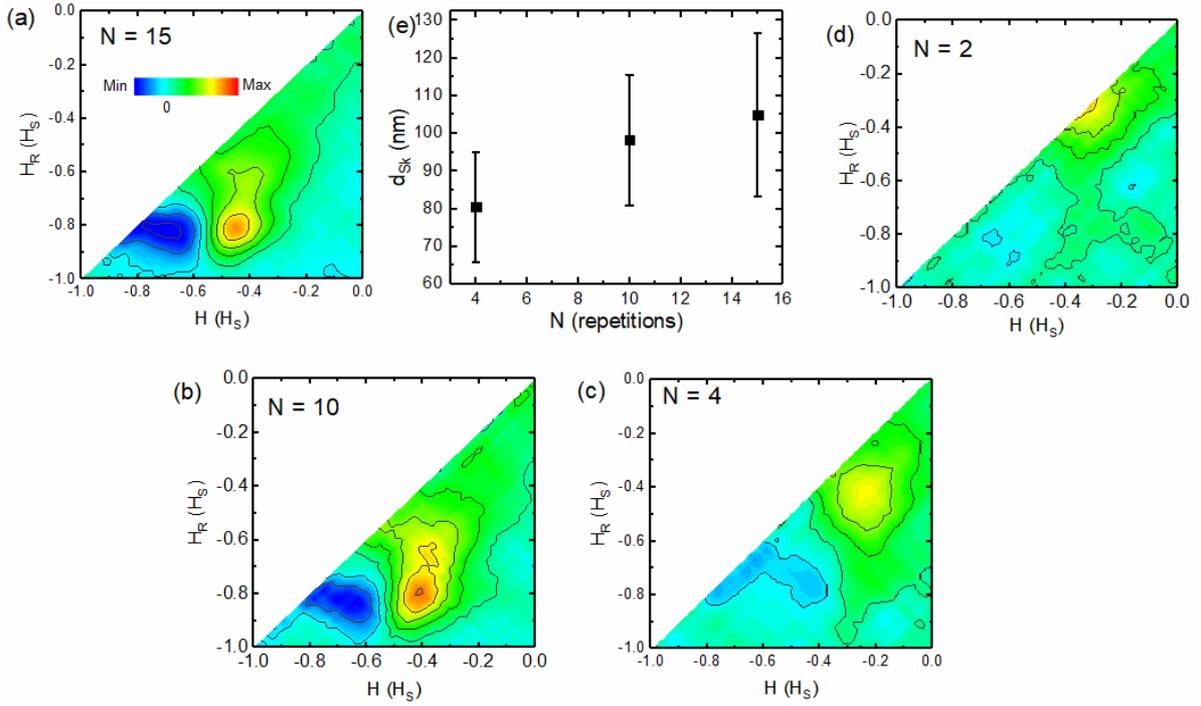

Figure 3: Evolution of FORC distributions with number of layer repetitions (N) in Pt/Co/MgO multilayers
(a)-(d) FORC diagrams for Pt/Co/MgO multilayers for N = 15, 10, 4, and 2, showing a gradual disappearance of the negative FORC valley as N decreases. Correspondingly, (e) the average skyrmion size decreases from ≈ 105 nm to ≈ 80 nm as N decreases from 15 to 4, reflecting a transition from a dipolar-dominant to DMI-dominant regime of skyrmion stability.

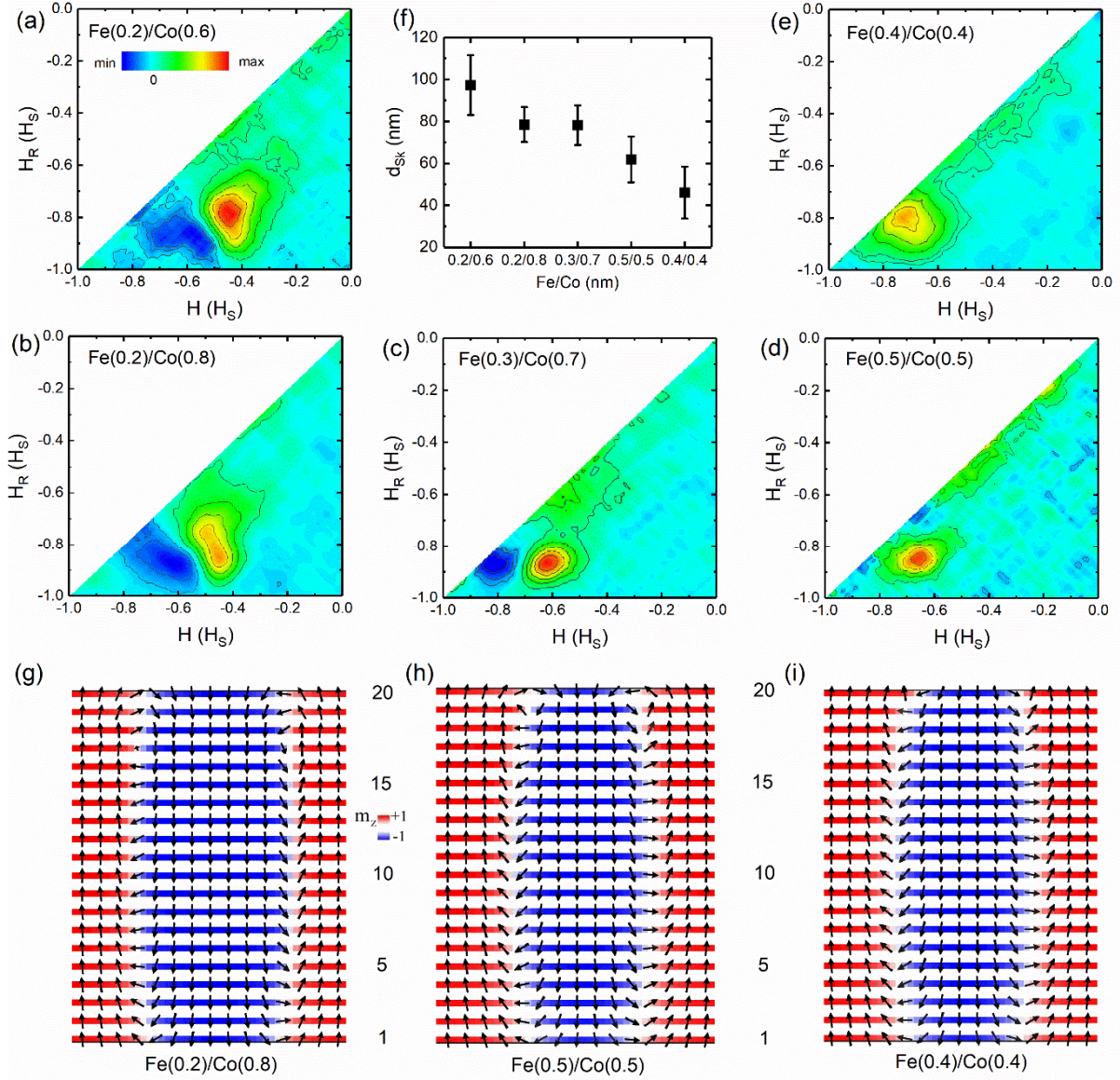

Figure 4: Evolution of FORC distributions with Fe/Co compositions in Fe(x)/Co(y) multilayers (a)-(e) FORC diagrams for various Fe(x)/Co(y) compositions: Fe(0.2)/Co(0.6), Fe(0.2)/Co(0.8), Fe(0.3)/Co(0.7), Fe(0.5)/Co(0.5), and Fe(0.4)/Co(0.4). Tuning the ferromagnetic compositions in this order brings the negative FORC valley closer to the diagonal edge: the negative valley eventually vanishes in Fe(0.4)/Co(0.4). (f) The skyrmion diameter correspondingly decreases from $\approx$ 97 nm to $\approx$ 46 nm across the samples, demonstrating an increasingly strong influence of the DMI which stabilizes pure Néel skyrmions in Fe(0.4)/Co(0.4). (g)-(i) Cross-sections of the skyrmion textures, obtained by micromagnetic simulations, for samples Fe(0.2)/Co(0.8), Fe(0.5)/Co(0.5), and Fe(0.4)/Co(0.4). For Fe(0.2)/Co(0.8), we used the following parameters: $M_s$=1238 kA/m, $K_{eff}$=287.01 kJ/m$^3$, $A$=10 pJ/m, $D$=1.5 mJ/m$^2$ and $H_{ext}$=190 mT. For Fe(0.5)/Co(0.5), we used: $M_s$=1010 kA/m, $K_{eff}$=59.05 kJ/m$^3$, $A$=10.3 pJ/m, $D$=1.9 mJ/m$^2$ and $H_{ext}$=200 mT.